\documentclass[prl,superscriptaddress,twocolumn]{revtex4}

\usepackage[latin1]{inputenc}
\usepackage{graphicx,psfrag}
\usepackage{bm}

\usepackage{txfonts}

\newcommand{\average}[1]{\left\langle{#1}\right\rangle}

\newcommand{\lij}{l_{ij}}

\newcommand{\pq}[1]{\left[{#1}\right]}

\begin{document}
\title{Mechanical unfolding and refolding pathways of ubiquitin}
\author{A. Imparato}
\affiliation{Dipartimento di Fisica and CNISM, INFN Sezione di Torino,   Politecnico di Torino,  c. Duca degli Abruzzi 24, 10129 Torino, Italy}
\author{A. Pelizzola}
\affiliation{Dipartimento di Fisica and CNISM, INFN Sezione di Torino,   Politecnico di Torino,  c. Duca degli Abruzzi 24, 10129 Torino, Italy}

\begin{abstract}
Mechanical unfolding and refolding of ubiquitin are studied by Monte
Carlo simulations of a G\={o} model with binary variables. The
exponential dependence of the time constants on the force is verified,
and folding/unfolding lengths are computed, with good agreement with
experimental results. Furthermore the model exhibits intermediate
kinetic states, as observed in experiments. Unfolding and refolding
pathways and intermediate states, obtained by tracing single secondary
structure elements, are consistent with simulations of previous
all--atom models and with the experimentally observed step sizes.
\end{abstract}
\maketitle

Understanding folding and unfolding pathways remains one of the major
challenges in protein science. Thermal and chemical
denaturation have been studied for decades, and in recent years new
experimental techniques, where single molecules are manipulated by
atomic force microscopy and optical tweezers, collectively referred to
as force spectroscopy \cite{rgo,CV_nsb,DR,Ober1,exp_fc1,exp_fc2},
have given a new perspective on this problem.
In typical experiments a protein is pulled by its
ends, and the unfolding and refolding processes are monitored by
measuring its end--to--end length as a function of time. In 
 force clamp experiments~\cite{exp_fc1,exp_fc2} the force is kept constant by
a feedback system. To unfold a molecule, the force is
suddenly increased from a small to a large value, and the reverse is
done to let it refold. Typically, unfolding  and refolding turn out to be two--state processes with a characteristic time which
depends exponentially on the force according to an Arrhenius--like
law,
but (un)folding
intermediates are also observed, as in ubiquitin. This is a small (76
aminoacids, pdb code 1ubq), highly stable protein which has been the
subject of many studies (see e.g.\
\cite{Jackson} and refs.\ therein). The native state contains an
$\alpha$--helix ($\alpha_1$: residues 23--34), a short $3_{10}$ helix
($\alpha_2$: 56--59) and a 5--stranded $\beta$--sheet forming tertiary
contacts with the $\alpha$--helix. The strands, in the order in which
they appear in the sheet, are $\beta_2$ (10--17), $\beta_1$ (1--7),
$\beta_5$ (64--72), $\beta_3$ (40--45) and $\beta_4$ (48--50).


Ubiquitin unfolding is characterized by a time constant which depends
exponentially on the force, and by a distance from the native to the
transition state equal to $x_u = 0.17$ nm \cite{exp_fc2}. The
unfolding transition is signalled by a 20 nm increase in the
end--to--end length. In a limited number of cases (5\% of the total)
however, a different unfolding pathway was observed, where the protein
unfolds in two steps, characterized by 8 and 12 nm length increases
respectively. This has been attributed to the existence of a partially
folded intermediate state containing $\alpha_1$, $\beta_1$ and
$\beta_2$~\cite{exp_fc2}, though in that experiment the structure of
such state has not been directly probed.

The refolding also exhibits a rich behaviour 
\cite{exp_fc1}, with an initial rapid elastic recoil, followed by an
intermediate state 
characterized by large length fluctuations, and a final transition to
the folded state. The folding time turns out to follow an Arrhenius--like
law, with a distance from the extended to the transition state
estimated as $x_f = 0.8$ nm \cite{exp_fc1}.


These experimental results have prompted a series of computational
studies~\cite{IMM,IM,LKH,Shakh,Cieplak1,Cieplak2} aimed at reproducing the general behaviour and elucidating the
details of the unfolding and refolding pathways and the nature of the
intermediate state. Irb\"ack and coworkers \cite{IMM,IM} 
suggest, on the basis of Monte Carlo (MC) simulations of an all--atom
model, that the most probable unfolding pathway corresponds to
$\beta_1 \beta_5 \rightarrow \beta_1 \beta_2 \rightarrow \beta_3
\beta_5 \rightarrow \beta_3 \beta_4 \rightarrow \alpha_1$, i.e., the
contacts between $\beta$--strands 1 and 5 are the first to yield,
followed by those between strands 1 and 2, and so on until finally the
$\alpha$--helix yields. Furthermore, in the typical unfolding
intermediate found in that paper, $\alpha_1$, $\beta_5$,
$\beta_3$ and $\beta_4$ are still folded, in marked contrast with
\cite{exp_fc2}. 

Li {\it et al.}~\cite{LKH} 
verified, using molecular dynamics (MD) simulations of a $C_\alpha$ G\={o}
model, that unfolding and refolding times depend exponentially on
the force, with $x_u =0.24$ nm and $x_f = 0.96$ nm.
They did not find an unfolding intermediate, and attributed this to
the lack of non--native interactions. They
distinguished three cases, where the force is applied to (a) both
termini, (b) N--terminus only, and (c) C--terminus only. In cases (a)
and (c) they obtained that the secondary structure elements (SSEs)
break in the order $\beta_1 \rightarrow \beta_2 \rightarrow \beta_5
\rightarrow \beta_3 \rightarrow \beta_4 \rightarrow \alpha$, while in
(b) they found $\beta_5 \rightarrow \beta_3 \rightarrow \beta_4
\rightarrow \beta_1 \rightarrow \beta_2 \rightarrow \alpha$.

Kleiner and Shakhnovich \cite{Shakh} 
found, using MC simulations of an all--atom G\={o} model, that 
unfolding starts with the separation of $\beta_1$, $\beta_2$
and $\beta_5$ from the rest of the structure and from each other,
followed by the separation of $\beta_3$ and $\beta_4$ from $\alpha_1$
and from each other, and finally by the unfolding of the helices. 
Their typical trajectory shows a plateau in the end--to--end
length, which they associate to an unstable intermediate where
$\beta_1$ and $\beta_2$ are unfolded and $\beta_5$ is about to
unfold.

Szymczak and Cieplak~\cite{Cieplak1,Cieplak2} observed, in MD
simulations of a $C_\alpha$ G\={o} model, that, during 
folding, the hairpin $\beta_1 \beta_2$ forms at the beginning
if both extremities are left free, while it forms at the end 
if the N-terminus is held fixed.


Motivated by the discrepancies between these results we have studied
the mechanical unfolding and 
refolding of ubiquitin by means of MC simulations of a simplified
G\={o} model with binary variables \cite{IPZ}. We have recently
developed this model as a generalization of a model originally
proposed by Wako and Sait\^o \cite{WS} in a 
purely thermodynamic version and subsequently reconsidered by Mu\~noz,
Eaton and coworkers \cite{ME1,MunozScience}, who used a
kinetic version of the model to analyze experimental results. In the
last few years it has been shown that the model equilibrium
thermodynamics can be solved exactly \cite{Ap1}, and that it
successfully describes the kinetics of protein folding
\cite{Amos,HenryEaton,ItohSasai1,Ap3,BPZ},
Our generalized model for protein mechanical unfolding
\cite{IPZ}, has been shown to exhibit the typical
response of  proteins to external loading, allowing one to estimate
the unfolding length of a titin domain, in excellent agreement with the
experimental value, and with a very
limited computational effort. 
In the model a protein made up of $N+1$ aminoacids is described as a
chain of $N$ peptide bonds: a
binary variable $m_k$ is associated to each bond and can live in two
states (native and unfolded, $m_k=1,0$). Given the $m_k$ values one
can identify native--like stretches (which can be as short as a single
aminoacid and as long as the whole chain) delimited by unfolded
bonds. A stretch goes from bond $i$ to $j$ if and only if $(1-m_i)
\prod_{i<k<j} m_k (1-m_j) = 1$, and to each stretch we associate (i) a
native length $l_{ij}$ taken from the pdb \cite{Length}, and (ii) an
orientational variable $\sigma_{ij}=\pm 1$, where $+1(-1)$ means
parallel (antiparallel) to the applied force \cite{Sigma}. Detailed
definitions have been given in \cite{IPZ}, where the parameter choice
is also described.
The energy scale for the ubiquitin model turns out to be
$\epsilon/k_B=156.6 $K, and the force unit is fixed so as  to match the experimental characteristic
unfolding force $f_u\simeq 35$  pN \cite{exp_fc1,Sigma}.
The unfolding and refolding kinetics are studied by MC simulations
with single--variable--flip Metropolis dynamics: the model time scale
$t_0$ corresponds to a single MC step, the temperature is taken to be
$T=300$ K.  In order to monitor the unfolding of SSEs, we define the
order parameter for each SSE as the fraction of its peptide bonds
in the native (folded) state: $m_{\beta_s} = \frac{1}{j_s-i_s}
\sum_{k=i_s}^{j_s-1} m_k$, with $s=1\dots5$, and where $\beta$--strand
$s$ includes aminoacids from $i_s$ to $j_s$, and similarly we define
$m_{\alpha_r}$, $r=1,2$ for the helices.

\begin{figure}[h]
\center
\includegraphics[width=8cm]{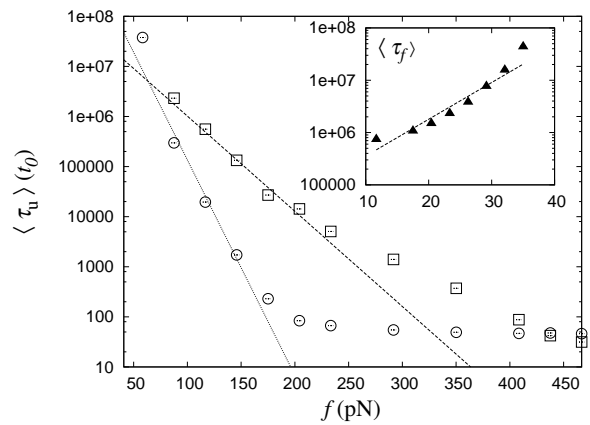}
\caption{Average unfolding time $\tau_u$ as a function of
the force $f$, applied to the whole molecule $(\square)$ and
to the portion of molecule spanning from the 48th to the 76th
aminoacid $(\bigcirc)$. Lines are exponential fits. Inset: Refolding
time as a function of the quenched force $f_1$, and with $f_0=100$
pN. The line is an exponential fit to the data.} 
\label{fc_fig}
\end{figure}

In the present work we consider the protein unfolding induced by a
force clamp \cite{exp_fc2}. 
Typically, in such experiment, the average unfolding time is given by
the Arrhenius' law $\average{\tau_u}= \tau_0 \exp\pq{-f x_u/(k_B T)}$,
where $f$ is the external force and $x_u$ the unfolding length.
We start
from the completely folded state, with $f=0$, and then
we apply a constant force $f$ and sample the unfolding time $\tau_u$ over
1000 trajectories.
In fig.~\ref{fc_fig} the mean unfolding time is plotted as a function
of the external force $f$. 
The force dependence is clearly exponential at small forces and
saturates at larger forces, as noticed in \cite{Cieplak1}. From a fit
of the data in fig.~\ref{fc_fig} to the Arrhenius' law we find
$x_u=1.8\pm 0.1$ \AA, in excellent agreement with the experimental
value found in \cite{exp_fc2}. The fit is performed in the same range
of forces considered in \cite{exp_fc2}, $50<f\lesssim 250$ pN.  In
ref.~\cite{CV_nsb}, the ubiquitin unfolding process has been shown to
depend on the pulling vector, relative to the structure. Following
that idea, we applied the force $f$ to the portion of molecule
spanning from the 48th aminoacid (Lys48) to the C terminus (aminoacids
48-76). We found bistability, signaling the unfolding transition, at
$f \sim 100$ pN, in reasonable agreement with the average unfolding
force of 85 pN measured in \cite{CV_nsb}.  We then measured
$\average{\tau_u}$ as a function of $f$ (fig.~\ref{fc_fig}). From the
fit we get $x'_u=4.1\pm 0.5$ \AA, which is larger than $x_u$ of the
whole molecule, signalling a softer structure, in agreement with
ref.~\cite{CV_nsb}. This value of $x'_u$ is slightly smaller than that
found in ~\cite{CV_nsb} (6.3 \AA), however in that work a different
experimental set up was used: the dynamic loading set-up. We plan to
address this discrepancy, as well as to give full description of the
unfolding of the ``48-76"--structure, in a future work.

\begin{figure}[h]
\center
\includegraphics[width=8cm]{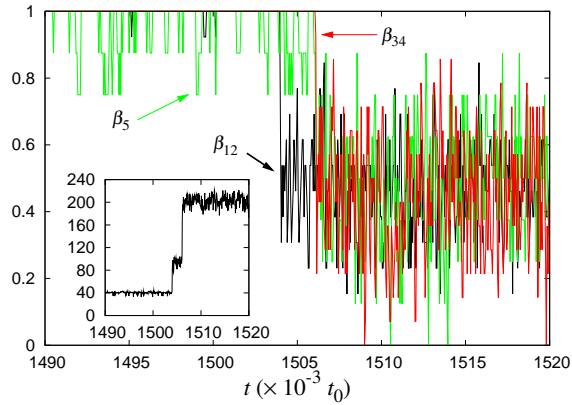}
\caption{(color online). Order parameters $m_{\beta_{12}}$ (black), $m_{\beta_{34}}$
  (red) and $m_{\beta_5}$ (green) as functions of time, with
  $f=100$ pN, for a typical trajectory. Inset: end--to--end length $L$ (in \AA) of the
  model molecule as a function of time.}
\label{Merge5}
\end{figure}

We probe the unfolding pathway by sampling the order parameters of the
single SSEs: $m_{\beta_s}$ and $m_{\alpha_r}$.
We checked that in the folded state, these order parameters take the
value 1 most of the time, with rare fluctuations where they take
smaller values. Let $t_{\beta_s}$ ($t_{\alpha_r}$) be the time at
which strand $\beta_s$ (helix $\alpha_r$) unfolds, defined as the time
at which the corresponding order parameter crosses a certain threshold
$m_u$ for the first time. Then, averaging over 1000 trajectories, we
compute the probability that, during unfolding at $f=100$ pN, a SSE
unfolds before another one. Results are reported in Table
\ref{Tab-Prec} and show clearly that the typical pathway starts with
the simultaneous unfolding of $\beta_1$ and $\beta_2$, followed by the
simultaneous unfolding of $\beta_3$, $\beta_4$ and $\beta_5$, and
finally by the helices unfolding.
\begin{table}
\begin{tabular}{|c|c|c|c|c|c|c|c|}
\hline
 & $\alpha_1$ & $\alpha_2$ & $\beta_2$ & $\beta_1$ & $\beta_5$ &
 $\beta_3$ & $\beta_4$ \cr 
\hline
$\alpha_1$ & $\times$ & 0.36 & 0.00 & 0.02 &0.05 & 0.07& 0.05 \cr
\hline
$\alpha_2$ & 0.64 & $\times$ & 0.02& 0.04 & 0.1&0.15 &0.11 \cr
\hline
$\beta_2$ & 1.00 & 0.98 & $\times$ & 0.57& 0.86 &0.945 &0.94 \cr
\hline
$\beta_1$ & 0.98&  0.96& 0.43 & $\times$ & 0.83&0.91 &0.90 \cr
\hline
$\beta_5$ & 0.95&  0.90& 0.14 & 0.17 & $\times$ & 0.64& 0.55\cr
\hline
$\beta_3$ & 0.93& 0.85 & 0.055& 0.09 &0.36 & $\times$ & 0.365\cr
\hline
$\beta_4$ & 0.95 &0.89 & 0.06 & 0.10 & 0.45 & 0.645& $\times$ \cr
\hline
\end{tabular}
\caption{Probability that the row--index SSE
  unfolds before the column--index one,  with 
  $f=100$ pN and   $m_u = 1/3$.}
\label{Tab-Prec} 
\end{table}
%
%
In Fig.\ \ref{Merge5} we plot the order parameters $m_{\beta_{12}}$
(for hairpin $\beta_1-\beta_2$, residues 1--17),
$m_{\beta_{34}}$ (residues 40--50) and $m_{\beta_5}$, in a time
interval containing the unfolding event of a typical
trajectory.
Three regions can be clearly distinguished. The top--left one
corresponds to the folded state, the rightmost one to the unfolded
state, and the central one to an intermediate state. In this intermediate
state  $\beta_1$ and $\beta_2$ are unfolded, $\beta_3$ and $\beta_4$ are
folded and $\beta_5$ fluctuates (as it already did in the folded state).
The presence of the intermediate state is also signalled by
a plateau in the end--to--end length, see inset of fig.~\ref{Merge5}.
This state appears in all but a few trajectories. Since
its lifetime is widely varying we argue that when we do not see it in
a trajectory, this is just due to the time resolution limit. The
unfolding pathway is then unique, exhibiting an 
intermediate state with a fluctuating lifetime. We have checked 
that the distribution of the time difference $t_{\beta_{34}}
- t_{\beta_{12}}$, which measures the time elapsed between the
unfolding of the hairpins $\beta_1-\beta_2$ and
$\beta_3-\beta_4$, 
exhibits a single peak (data not shown), while in the case of two different pathways one would expect a two--peaked distribution.


Our pathway is consistent with the all--atom 
models \cite{IMM,Shakh} and the $C_\alpha$ G\={o}
model by Li et al. \cite{LKH} (except for the case in which the force
is applied to the N--terminus only). In the experimental
reference~\cite{exp_fc2} the intermediate was attributed to the
partial unfolding of the strands $\beta_1$ and $\beta_2$ and of the
$\alpha$--helix, although this conclusion was based only on the length
of the single strands, their unfolding trajectories being not
experimentally accessible.  On the contrary, in ref.~\cite{IMM} the
intermediate state were identified to be composed by 
$\beta_3$, $\beta_4$ and $\beta_5$ and the $\alpha$--helix, as found
in the present work, though $\beta_5$ is fluctuating here. 
Also in ref.~\cite{Shakh} it was found that in the intermediate state
$\beta_1$ and $\beta_2$ are unfolded, 
while $\beta_5$ separates from $\beta_3$ along the plateau  which
characterizes the intermediate. 
The apparent discrepancy between theoretical and experimental
scenarios for the intermediate state can be reconciled if one
considers that the only direct information which comes from the
experiments is that in the first unfolding step a portion of the size
of 28 aminoacids unfolds \cite{exp_fc2}. The hairpin $\beta_1-\beta_2$, plus the
loop which connects it to the $\alpha$--helix, measures 22 aminoacids,
and the remaining 6 aminoacids can be attributed to the
fluctuations of the strand $\beta_5$, that we observe in the intermediate state. 

\begin{figure}[h]
\center
\includegraphics[width=8cm]{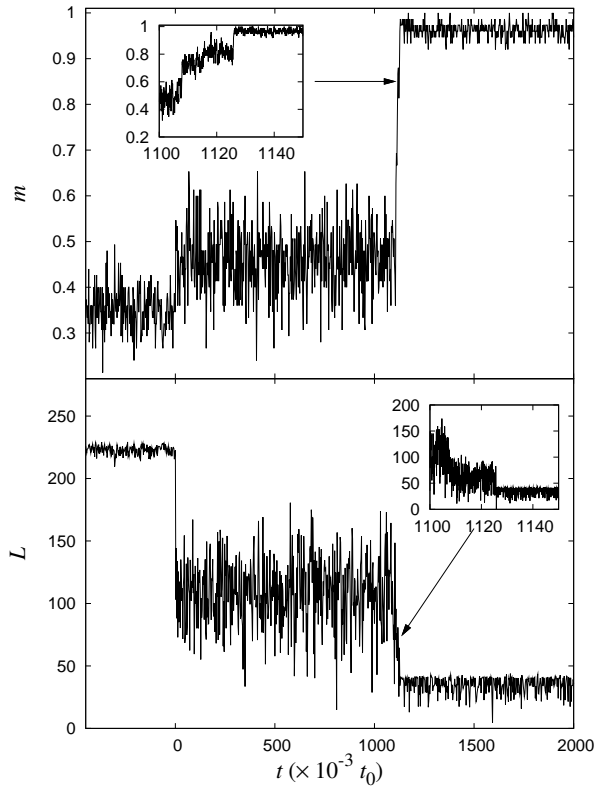}
\caption{Refolding under constant force: molecular order parameter $m$ and length $L$ as  functions of the time, $f_0=232$ pN, $f_1=23.2$ pN.}
\label{ref_fig_f}
\end{figure}

We now turn to the analysis of refolding, by 
considering a protein which is
initially completely unfolded, at equilibrium with a large
force $f_0$. Then, at $t=0$, the stretching force is quenched to a low
value $f_1$, and the folding trajectory of the protein is monitored.  In
fig.~\ref{ref_fig_f} a typical trajectory is shown: the order
parameter $m$ (fraction of native peptide bonds \cite{IPZ} ) and end--to--end 
length $L$ are plotted as functions of time.
It is worth noting that, after the quench at $t=0$, we observe a
fast elastic recoil, where the length jumps to a
value $L\simeq 100$ \AA, while the order parameter $m$ increases more
gradually, indicating that the molecule is not yet structured. After
this recoil, the length and the order
parameter exhibit large fluctuations. This stage is followed by
another abrupt contraction in the length, where the protein
reaches its equilibrium length for the small force applied. This is
accompanied by a marked increase in the order parameter, which shows
that the molecule is now fully folded. This behaviour is the same that
was found by Fernandez and Li \cite{exp_fc1} in their experimental
observation of refolding of ubiquitin under force quenching. It is
worth stressing that in a subset of trajectories, they found that the
last transition can be split in two stages, however the order of
refolding of the SSEs during these two stages could not be sampled.
\begin{table}
\begin{tabular}{|c|c|c|c|c|c|c|c|}
\hline
 & $\alpha_1$ & $\alpha_2$ & $\beta_2$ & $\beta_1$ & $\beta_5$ &
 $\beta_3$ & $\beta_4$ \cr 
\hline
$\alpha_1$ & $\times$ & 0.975 & 1.00 & .99 & 0.99 &0.98 &0.98 \cr
\hline
$\alpha_2$ &0.025 & $\times$ &1.00 &1.00 & 1.00& 0.975 & 0.975\cr
\hline
$\beta_2$ & 0.00 & 0.00 & $\times$ &0.54 &0.37 &0.14 &0.14 \cr
\hline
$\beta_1$ & 0.01 &  0.00& 0.56 & $\times$ & 0.365& 0.13& 0.13\cr
\hline
$\beta_5$ &0.01  &  0.00& 0.63 & 0.635& $\times$ &0.09 & 0.085\cr
\hline
$\beta_3$ &0.02 & 0.025 & 0.86 & 0.87 & 0.91 & $\times$ & 0.97\cr
\hline
$\beta_4$ & 0.02& 0.025 & 0.86 &0.87  & 0.915 & 0.03& $\times$ \cr
\hline
\end{tabular}
\caption{Probability that the row--index SSE refolds before the
  column--index one, with $f_0=232$ pN, $f_1=23.2$ pN.}
\label{Tab-Prec-ref} 
\end{table}
We find that the refolding pathway is similar to the reverse unfolding one, see table \ref{Tab-Prec-ref}.
 The helices and
the strands $\beta_3$, $\beta_4$  are the first SSEs to
refold,  $\beta_1$ and $\beta_2$  fold at the final stage of the process, while 
$\beta_5$ folds randomly between $\beta_3-\beta_4$ and $\beta_1-\beta_2$: thus
in addition to the fast elastic recoil the model exhibits also a refolding
intermediate (see the insets in Fig.\ \ref{ref_fig_f}).
It is tempting to associate this intermediate to the two stages observed
in ref.~\cite{exp_fc1}
in the last transition.
Finally, we observe that the average refolding time depends also
exponentially on the force $\ln <\tau_f>\sim f x_f$, as shown in the inset of Fig.\ \ref{fc_fig}, and the
corresponding folding length is $x_f = 6.7\pm 0.8$ \AA, in reasonable agreement with the
experimental value  $x_f = 8$ \AA~\cite{exp_fc1} and with the value obtained by Li et
al. \cite{LKH}.


%

In conclusion, we have shown that a simple G\={o} model with binary
variables can account for the main features observed in the
mechanical unfolding and refolding of ubiquitin.  This model is, to
the best of our knowledge, the simplest one with sufficient details to
allow the study of specific molecules.  
We believe that this model may be a useful
tool to investigate the interplay between the protein native
structures and their unfolding and refolding pathways.
Finally, we want to stress that the ubiquitin refolding cannot
be investigated by all-atom simulations because of the huge computation time 
required.


\begin{thebibliography}{}
\bibitem{Length} In our model the native length $\lij$ is independent
  of the force. This simplification leads to overall lengths slightly
  smaller than the experimental ones, as the extensibility of the
  native stretches is not taken into account.
\bibitem{Sigma} Our assumption on orientations
  corresponds to a reduction in configuration space and hence in the
  entropy of unfolded bonds: as a consequence we
  have to rescale the ``natural'' force unit $\epsilon/(.1 \mathrm
  {nm})$ by a factor 5.4, in such a way that the force $f_u$ where the
  molecule elongation is half of its maximum value, corresponds to the
  experimental unfolding force $f_u=35$ pN
  \cite{exp_fc1}.
\end{thebibliography}
\end{document}